\documentclass[english,reprint,amsmath,amssymb,aps]{revtex4-2}
\usepackage[latin9]{inputenc}
\usepackage{mathtools}
\usepackage{amsmath}
\usepackage{graphicx}

\makeatletter
\usepackage{bm}
\usepackage{xcolor}
\usepackage[mathscr]{euscript}

\usepackage{babel}

\makeatother

\usepackage{babel}
\begin{document}
\title{Controlling molecular dynamics by exciting atoms in a cavity }
\author{Andr\'as Csehi$^{1}$, \'Agnes Vib\'ok$^{1,2}$, Lorenz S. Cederbaum$^{3}$,
and G\'abor J. Hal\'asz$^{4\!\!\!}$}
\affiliation{$^{1}$Department of Theoretical Physics, Faculty of Science and Technology,
University of Debrecen, H-4002 Debrecen, PO Box 400, Hungary}
\affiliation{$^{2}$ELI-ALPS, ELI-HU Non-Profit Ltd., H-6720 Szeged, Dugonics t\'er
13, Hungary}
\affiliation{$^{3}$Theoretical Chemistry, Institute of Physical Chemistry, Heidelberg
University, Im Neuenheimer Feld 229, 69120 Heidelberg, Germany}
\affiliation{$^{4}$Department of Information Technology, Faculty of Informatics,
University of Debrecen, H-4002 Debrecen, PO Box 400, Hungary}
\begin{abstract}
Placing an atom and a molecule in a cavity opens the door to initialize
molecular dynamics by exciting a level of the atom. This approach
enlarges the range of choosing the light source to trigger molecular
dynamics substantially. The interplay of the atomic, molecular and
photonic populations gives rise to rich dynamics. The cavity photon
plays the role of a mediator between the atom and the molecule and
it is found that the photonic population is rather low throughout
and its evolution follows that of the molecule. Cavities are known
to be subject to losses. In spite of the losses it is demonstrated
that the presence of the atom gives rise to a long-lived dynamics
which should be of relevance for experimental investigations. The
presence of more atoms and molecules is expected to further enrich
the dynamics.

\vspace*{-8mm}~.
\end{abstract}
\maketitle

Fabry-Perot and plasmonic nano-cavities have been found to be useful
tools for manipulating atomic and molecular properties. The confined
photonic mode of the cavity can resonantly couple to the transition
dipole of the matter giving rise to mixed polaritonic states carrying
both photonic and excitonic features. In the last decade several experimental
and theoretical works have demonstrated that these hybrid vibronic
(visible) and vibrational (infrared) polaritons can dramatically modify
the different properties of the molecular systems \citep{Hutchison20121592,cavity_Zhong_AngChem_2016,Ebbesen20162403,Galego2015,Flick20173026,Herrera2017,Ribeiro20186325,Ruggenthaler2018,Feist2018205,Herrera2020,Garcia-Vidal2021,Feist2022}.
Among others these can enhance or suppress photochemical reaction
rate \citep{Galego2016,Munkhbat2018,Sun2022}, control of chemical
reactions by varying the properties of the quantized electric field,
\citep{Li2021,Schfer2022,George2023} enhance charge and long range
energy-transfer processes, \citep{Du2018,Reitz2018,Perez2019,Schfer2019,Mandal2020,Xiang2020,TorresSnchez2021,Cederbaum2021,Chen2022,Liu2020}
and increase non-adiabatic effects in molecules \citep{Kowalewski20162050,Fregoni2018,Szidarovszky20186215,Csehi2019,Csehi2019a,Ulusoy20198832,Perez2019,Triana2019,Gu20201290,Szidarovszky2020,Gu2020a,D1CP00943E,Cederbaum2021a,Fabri2021a,Badank2021,Fbri2022}.

One way to successfully modify and control molecular properties is
to reach the strong coupling regime such that the impact of the coupling
strength between light and matter excitations is larger than that
of the photonic and material losses. This can be reached by letting
a large number of molecules interact with the cavity mode \citep{Herrera2016,Galego2017,Groenhof2018,Groenhof2019}.
As for a single molecule, the situation is more complicated \citep{Zengin2015,Chikkaraddy2016127,TorresSnchez2021}.
Here to achieve sufficiently strong coupling nowadays and handling
correctly the physical situation, subwavelength size plasmonic nano-cavity
is needed. The latter however, may posses a very lossy behavior which
should be treated properly by the employed numerical simulations \citep{Khurgin2015,Felicetti20208810,Ulusoy2020a,Davidsson2020234304,Antoniou2020}. 

In the present work, we consider a complex quantum system consisting
of three different types of entities; a cavity radiation mode, a molecule
and an atom. Such systems have been investigated before in a different
context \citep{Ulusoy20198832,Szidarovszky2020,Davidsson2020}. In
order to be able to initialize any kind of dynamics in a single molecule
placed in a cavity, energy must be supplied to the system typically
done by using a laser. In our present quantum system the energy of
the pumping laser can transferred to the molecule with the help of
the atom. A particular advantage follows from that atoms possess many
discrete electronic energy levels and one can easily find a suitable
pair whose energy difference is close to resonance with the cavity
photon as well as with the desired energy transition of the molecule
at the geometry of choice, for example, in the vicinity of the molecule\textquoteright s
Franck-Condon (FC) geometry. The atomic level can be excited resonantly
by the laser allowing, for example, to use a rather weak laser pulse
which does not affect the molecule directly. The atom acts as an energy
converter, providing much flexibility in the choice of the laser source
even far from the molecular resonance range. The energy pumped into
the system by means of an atom is then transmitted to the molecule
via the cavity photon.

We study the time evolution of the energy exchange process taking
place between the three different components when the molecule is
able to communicate with the atom only via the cavity photon. We focus
on investigating how the energy travels back and forth between the
different components of our atom-cavity-molecule hybrid system. The
dynamics of our mixed quantum system is discussed for both photon
lossless and lossy situations. It will be demonstrated that despite
of the very short lifetime of a single photon in a plasmonic nano-cavity,
the back and forth energy flow between the three objects persists
for an extremely long time. The oscillations of the populations of
the different entities of the hybrid system live long and keep the
dynamics alive for quite a while making this system amenable to experimental
investigations even at times much longer than the lifetime of the
cavity photon. The lifetime of a nano-resonator is very short, but,
on the other hand, if three different participants share the energy,
and here the emphasis is on different, then the dynamics can last
for a very long time due to the role of cavity mode\textquoteright s
as a mediator. This effect is strictly not linear, but depends on
certain circumstances. 

\begin{figure}
\includegraphics[width=1\columnwidth]{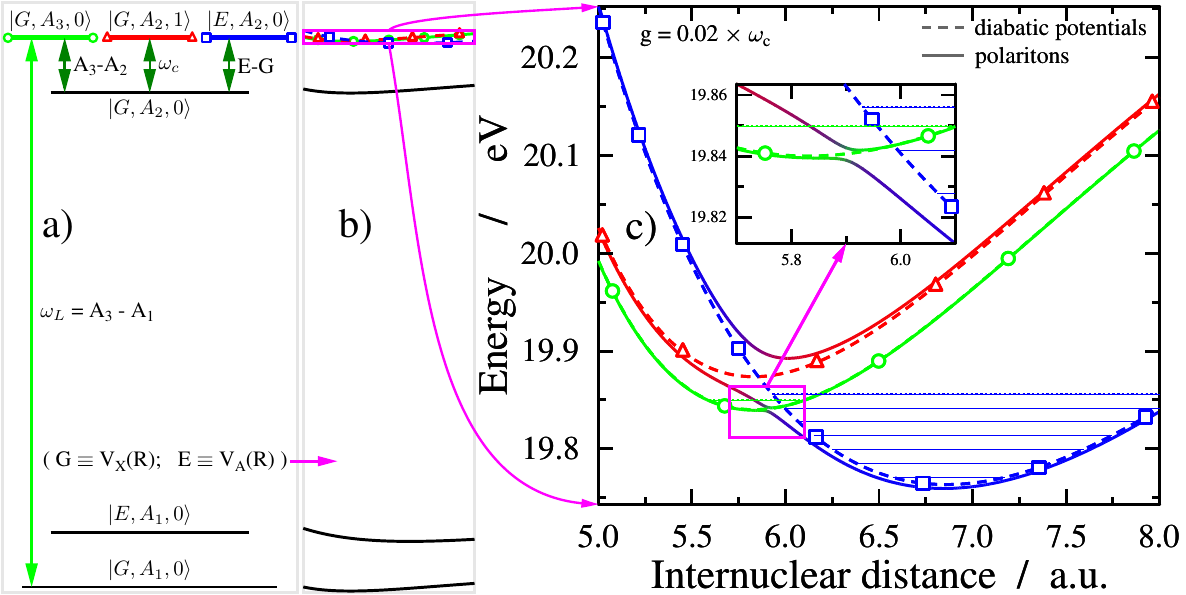}

\caption{\label{Fig.1}Scheme explaining the energetics of our show case example
of the three players atom-cavity-molecule ($\mathrm{Ne}$-cavity-$\mathrm{Na_{2}}$)
system. Panel a) shows the diabatic energy states of the studied system
without the molecular vibration. The ground state of the system indicated
by $\mathrm{\left|G,A_{1},0\right\rangle }$, where $\mathrm{G}$,
$\mathrm{A_{1}}$ and $\mathrm{0}$ are the molecular, atomic and
cavity photon states, is excited by the laser (green vertical line)
to the $\mathrm{\left|G,A_{3},0\right\rangle }$ state. This state
is essentially degenerate with the$\mathrm{\left|G,A_{2},1\right\rangle }$
and $\mathrm{\left|E,A_{2},0\right\rangle }$ states, where E stands
for the excited molecular state. Panel b) shows the diabatic potentials
for the molecular vibrations. Panel c) provides the detailed structure
of the first-excited manifold and the inset enlarges the region of
the narrow avoided crossing. The colors correspond to those shown
in panel a).\vspace*{-5mm}}
\end{figure}

As a realistic show case example, we consider a $\mathrm{Na_{2}}$
molecule and a Ne atom. Fig.\ref{Fig.1} shows the Ne atom in a three-state
representation (with energy levels $\mathrm{A_{1},A_{2}\,and\,A_{3}}$)
and the $\mathrm{Na_{2}}$ molecule in a two-state representation
(with potential energies $\mathrm{V_{X}(R)=X^{1}\varSigma_{g}^{+};V_{A}(R)=A^{1}\varSigma_{u}^{+}}$)
interacting with the quantized electromagnetic mode of a cavity (with
photon frequency $\mathrm{\omega_{c}}$). The electronic structure
data and the actual parameter values are given in the Supplementary
Information (SI). The atom and molecule are coupled to the cavity
mode and to the external electric laser field $\mathrm{E(t)}$, but
not directly to each other. To describe the nearly degenerate subspase
shown in Fig. \ref{Fig.1} we employ the following total Hamiltonian
of the atom-cavity-molecule system (atomic units are used throughout):
\begin{equation}
\begin{split}\mathrm{\hat{H}(t)}= & \mathrm{-\frac{1}{2M}\frac{\partial^{2}}{\partial R^{2}}{\bf 1}+\mathrm{\big(-\frac{1}{2}\frac{\partial^{2}}{\partial x^{2}}+\frac{1}{2}\omega_{c}^{2}x^{2}\big){\bf 1}+\hat{V}(t)}}\\
\hat{V}(t)= & \begin{pmatrix}\mathrm{V_{X}}+\mathrm{A_{1}} & 0 & \mathrm{c_{13}(t)} & 0\\
0 & \mathrm{V_{X}+A_{2}} & \mathrm{c_{23}}(t) & \mathrm{C_{XA}(t)}\\
\mathrm{c_{13}(t)} & \mathrm{c_{23}}(t) & \mathrm{V_{X}+A_{3}} & 0\\
0 & \mathrm{C_{XA}}(t) & 0 & \mathrm{V_{A}+A_{2}}
\end{pmatrix}
\end{split}
\label{eq:hfull4}
\end{equation}
where ${\bf 1}$ symbolizes the 4 $\times$ 4 unit matrix, M is the
reduced mass and R is the molecular internuclear coordinate. In eq.
(\ref{eq:hfull4}), the cavity is treated as a harmonic oscillator
using the photon displacement coordinate \citep{Vendrell2018a,Csehi2019a}.
The $\mathrm{V_{11}}$ element of eq. (\ref{eq:hfull4}) describes
the electronic ground state of the system. $\mathrm{V_{22}}$ and
$\mathrm{V_{33}}$ describe the atomic excitations while $\mathrm{V_{44}}$
the relevant combined atomic and molecular excitation, see Fig. \ref{Fig.1}.
In the case of a lossy cavity, an additional $\mathrm{-\frac{i}{2}\kappa}$
imaginary term is acting on the photonic excitation. It accounts for
the loss of the cavity photon by absorbing the photonic wave function
whenever the system is in the photonic state. $\mathrm{\kappa}$ is
the cavity decay rate which is inversely proportional $\mathrm{\kappa=\frac{1}{\tau}}$
to a lifetime $\tau$ of the cavity photon and the cavity quality
factor $Q=\frac{\omega_{C}}{\kappa}$, as described in detail in \citep{Felicetti20208810,Ulusoy2020a,Fabri2024}.
Finally, the $c_{k\ell}$ and $C_{XA}$ terms account for the coupling
of the atom and of the molecule, respectively to the electric field
of the laser and to the cavity, and they have the explicit form in
the dipole approximation \citep{Vendrell2018a,Csehi2019a}
\begin{align}
\mathrm{c_{k\ell}(t)=} & \mathrm{-d_{A_{k}A_{\ell}}E(t)+gd_{A_{k}A_{\ell}}\sqrt{2\omega_{c}}x}\quad(\mathrm{k\neq\ell=1,2,3})\nonumber \\
C_{XA}(t)= & -\mu(R)E(t)+g\mu(R)\sqrt{2\omega_{c}}x.\label{eq:TDM-1}\\
E(t)= & E_{0}\cdot sin^{2}(\pi t/T)\cdot cos(\omega_{L}t)\nonumber 
\end{align}
Here $\mathrm{d_{A_{k}A_{\ell}}}$ and $\mathrm{\mu(R)}$ are the
atomic and molecular transition dipoles, respectively, $\mathrm{g}$
is the cavity coupling strength. In the laser pump pulse used to initiate
the dynamics $T$ is the pulse duration, $\mathrm{E_{0}}$ is the
electric field amplitude and $\mathrm{\omega_{L}}$ is the central
frequency. 
The dynamics is initiated by pumping the $\mathrm{A_{3}}$ state of
the atom setting $\mathrm{T=100}$ fs, $\mathrm{I_{0}=1\times10^{12}W/cm^{2}}$
and $\mathrm{\omega_{L}\thickapprox20.57\,eV}$. 

\begin{figure}
\includegraphics[width=0.49\columnwidth]{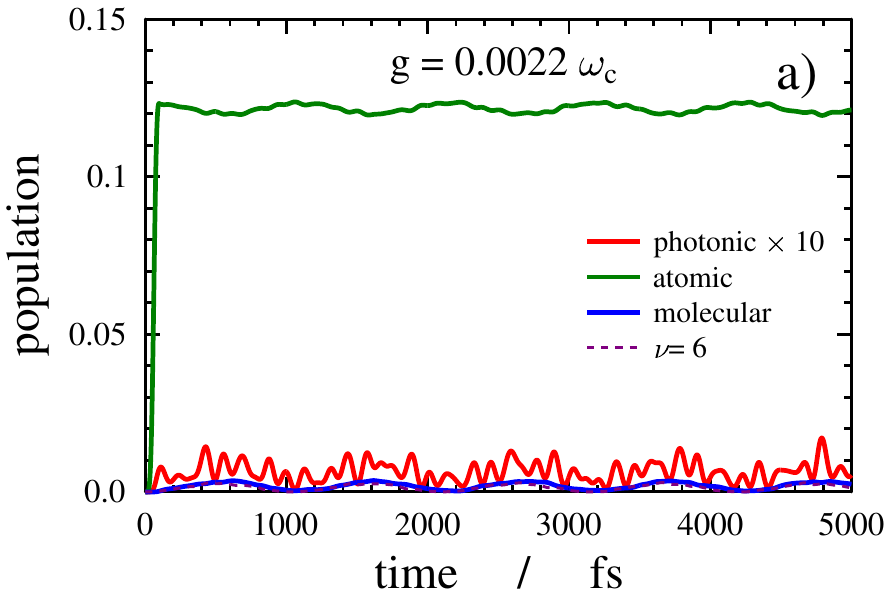}\includegraphics[width=0.49\columnwidth]{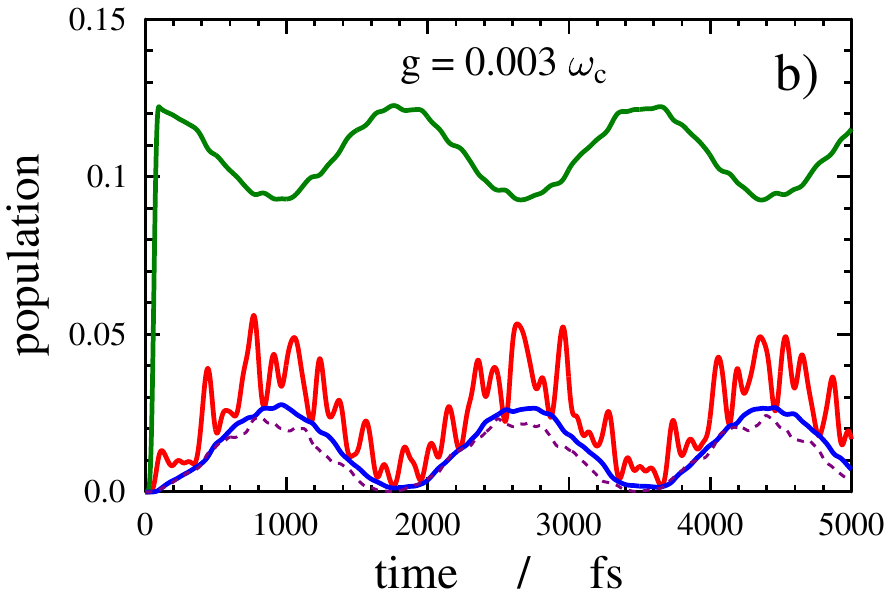}

\includegraphics[width=0.49\columnwidth]{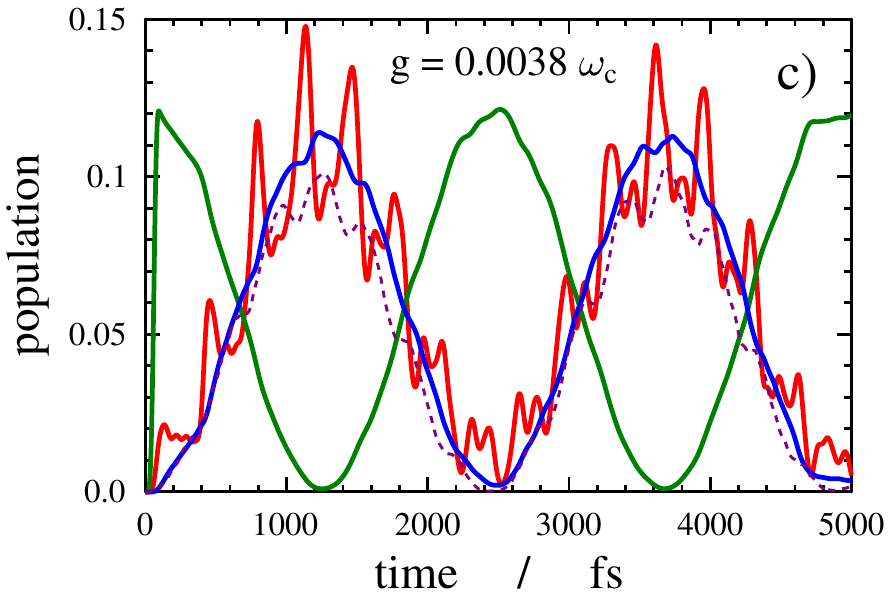}\includegraphics[width=0.49\columnwidth]{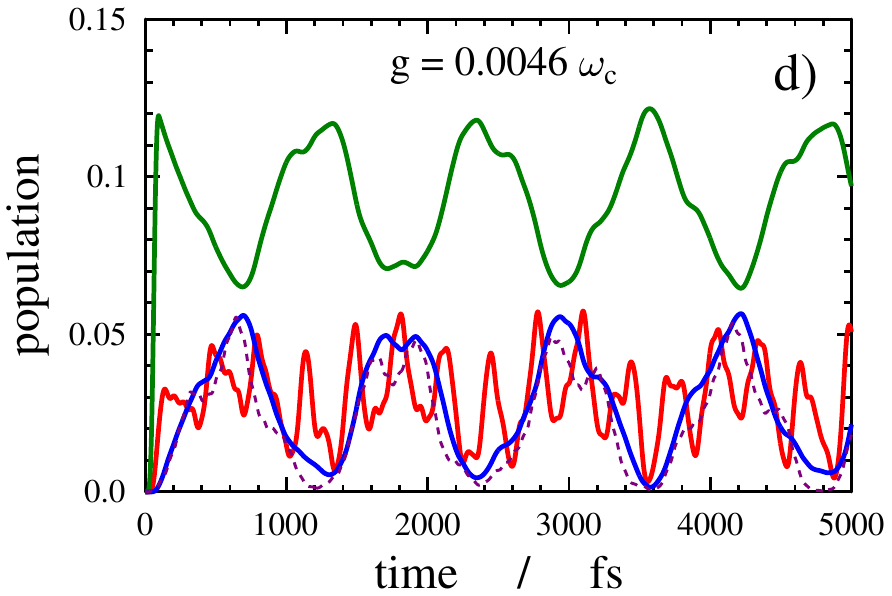}

\caption{\label{Fig.2}Population dynamics of the excited atom and molecule
and photon. Selected cavity coupling strength values are applied in
panels a)-d) for a lossless cavity. Note that the population of the
cavity photon is magnified by a factor of 10. The population of the
$\mathrm{\nu=6}$ vibrational level of the diabatic state $\mathrm{\left|E,A_{2},0\right\rangle }$
is also shown by a dashed line. It is seen to be rather close to that
of the molecular population. See text for discussion.\vspace*{-5mm}}
\end{figure}
\begin{figure}
\includegraphics[width=0.49\columnwidth]{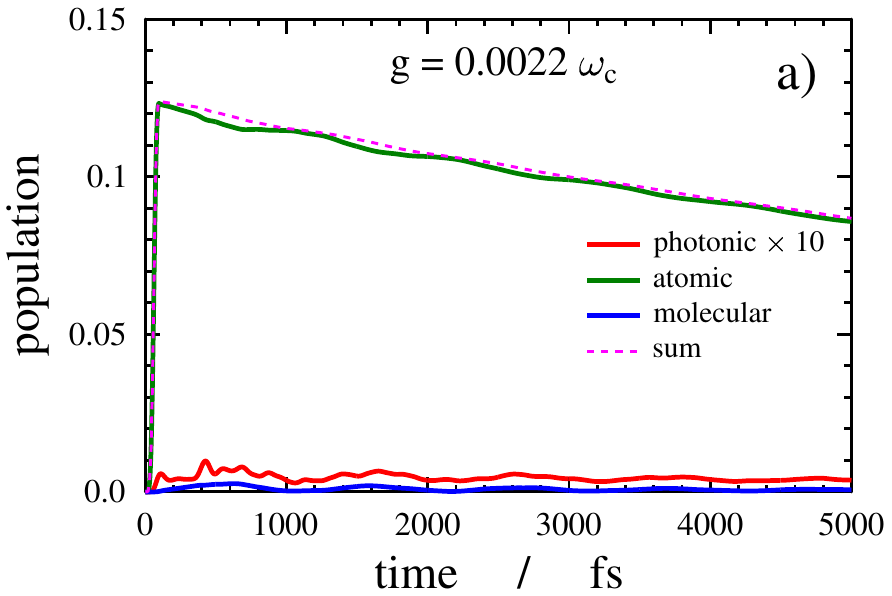}\includegraphics[width=0.49\columnwidth]{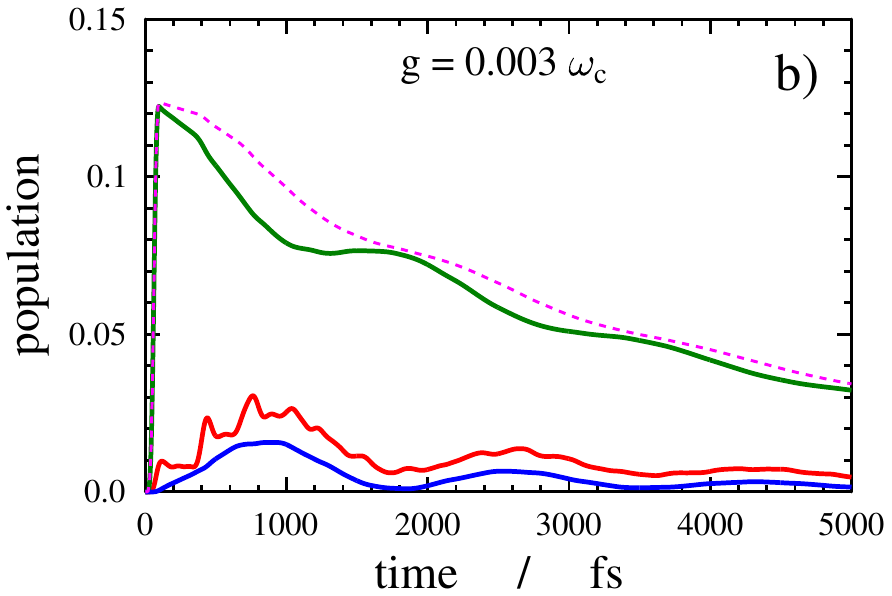}

\includegraphics[width=0.49\columnwidth]{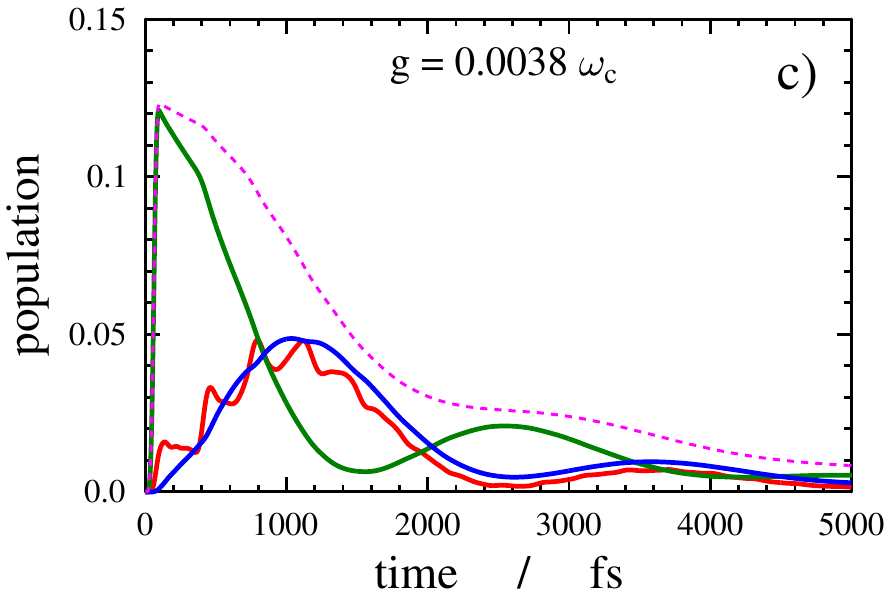}\includegraphics[width=0.49\columnwidth]{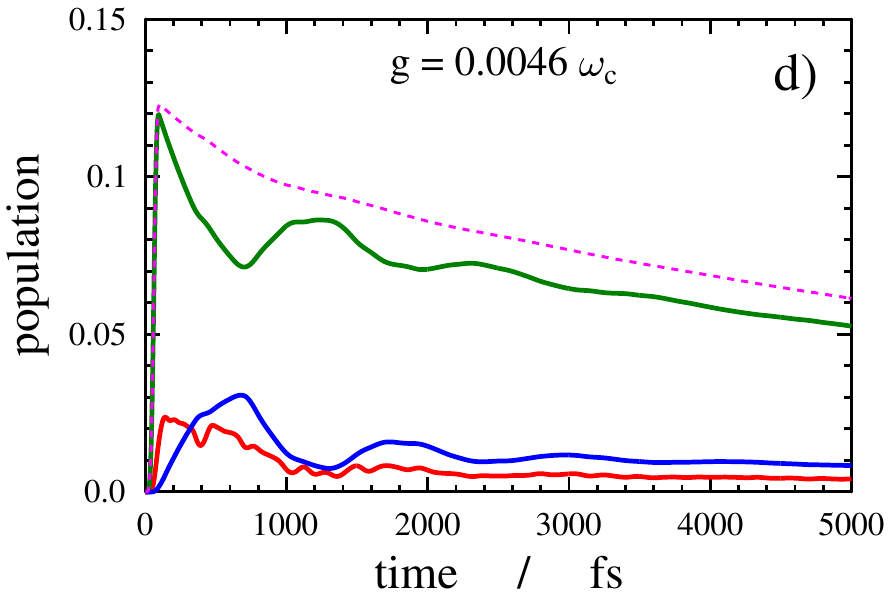}

\caption{\label{Fig.3}Population dynamics of the excited atom and molecule
and photon for a cavity with losses.The coupling strengths applied
in panels a)-d) are the same as in Figure \ref{Fig.2}. The population
of the cavity photon is magnified by a factor of 10. The cavity losses
are for $\mathrm{\mathrm{\kappa=0.0004\,au}}$ implying a photon lifetime
of $\mathrm{\tau=\mathrm{61\,fs}}$. This value is compatible with
the lifetime of plasmonic nano-cavities \citep{Feist2022}.\vspace*{-3mm}}
\end{figure}

\begin{figure}
\includegraphics[width=0.49\columnwidth]{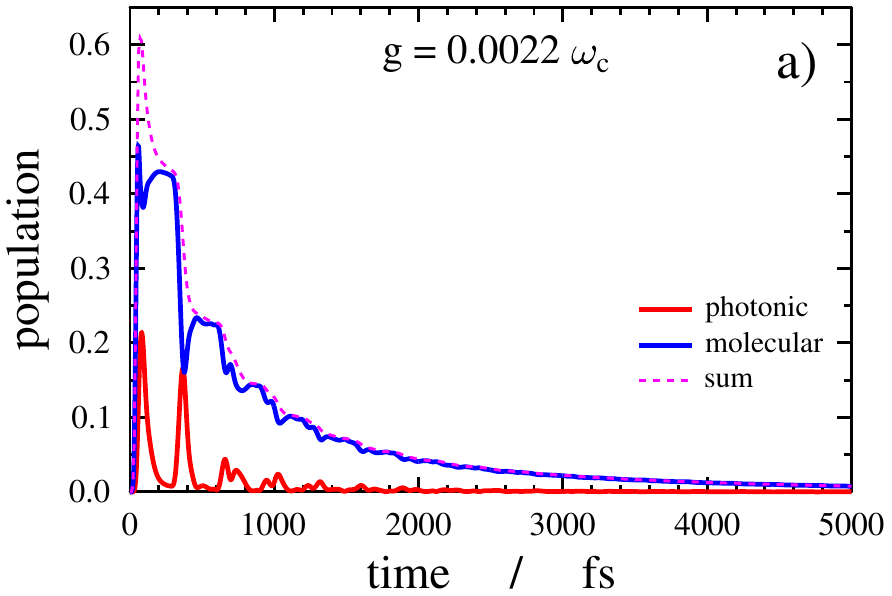}\includegraphics[width=0.49\columnwidth]{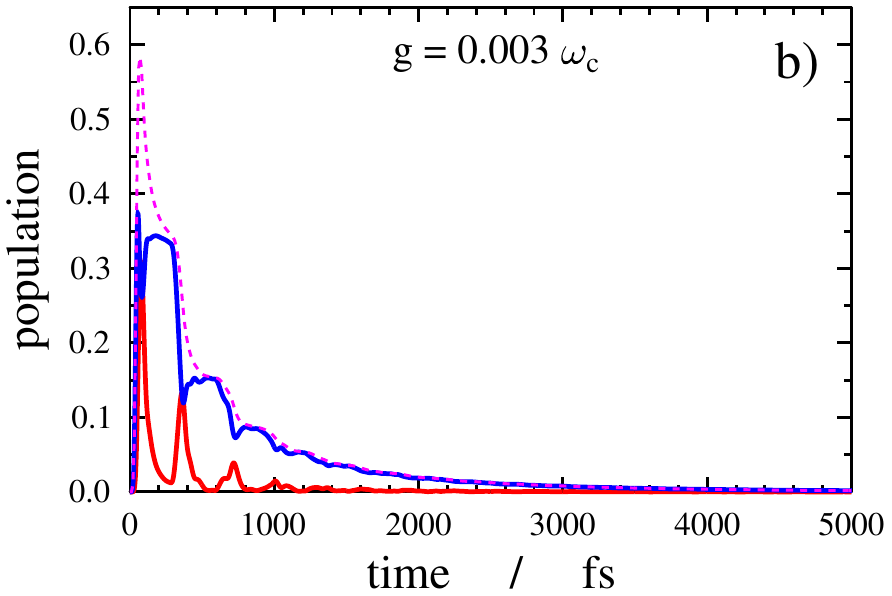}

\includegraphics[width=0.49\columnwidth]{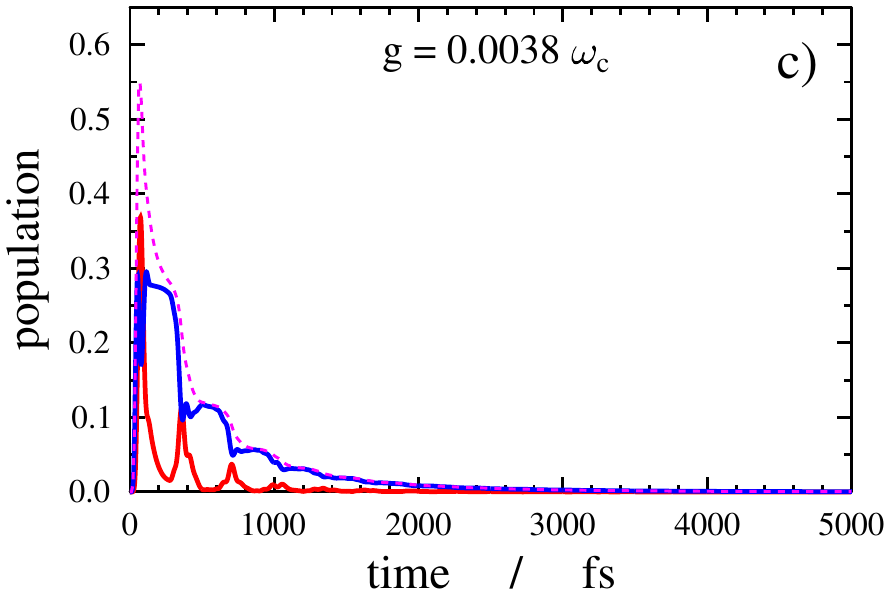}\includegraphics[width=0.49\columnwidth]{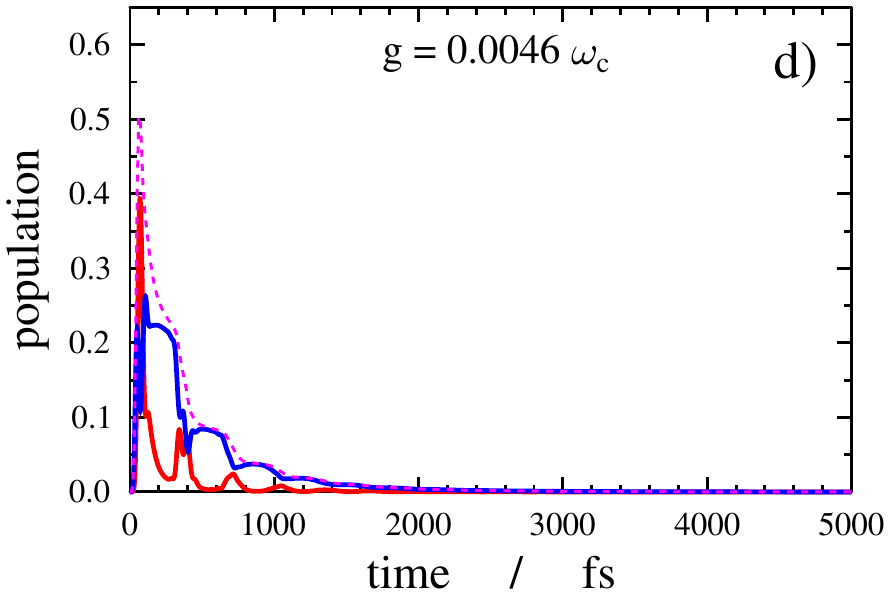}

\caption{\label{Fig.4}Population dynamics of the excited molecule and cavity
photon for a cavity with losses. In contrast to Fig. \ref{Fig.3},
the atom has been removed and the laser frequency has been changed
to pump the excited molecular state directly in order to study the
impact of the atom. The laser frequency is now $\mathrm{1.968\,eV}$.
All the other quantities employed are as in Fig. \ref{Fig.3}.\vspace*{-5mm}}
\end{figure}

The MCTDH (multiconfiguration time-dependent Hartree) method \citep{Manthe1992,Beck2000}
is applied to solve the time-dependent Schrödinger equation using
the Hamiltonian (\ref{eq:hfull4}). The details of the calculations
are provided in the SI. The convergence of the results has been checked.

To follow the dynamics of the system, we computed and investigated
the atomic and molecular populations in their respective excited state
as well as the photonic population as a function of time. These relevant
quantities follow from the populations of the various four components
$\mathrm{\Psi_{j}}$ of the four dimensional MCTDH total wave function
which read:
\begin{equation}
\mathrm{p_{j}(t)=\langle\Psi_{j}(R,a,x,t)|\Psi_{j}(R,a,x,t)}\rangle\quad(\mathrm{j=1,2,3,4}).\label{eq4}
\end{equation}

Fig. \ref{Fig.2} shows how the population migrates between the different
players as a function of time for four coupling strengths g of the
cavity. As indicated in Fig. \ref{Fig.1}a, the system is initially
in its ground state $\mathrm{\left|G,A_{1},0\right\rangle }$ and
the external laser field takes it to the state $\left|\mathrm{G,A_{3},0}\right\rangle $
where the atom is excited. The pulse duration is $\mathrm{100\,fs}$
(see below eq. (\ref{eq:hfull4})) explaining the sharp rise of the
atomic population on the long time scale of Fig. \ref{Fig.2}. The
cavity mode employed is chosen to be resonant with the energy difference
between the ground and excited molecular states at the FC point ($\mathrm{\omega_{c}=1.968eV}$)
and nearly resonant with the energy difference ($\mathrm{A_{3}-A_{2}=1.9343eV}$)
of the atom. By pumping the $\mathrm{A_{3}}$ level of the atom with
the laser, it now becomes possible to transfer population from the
atom to the molecule via the coupling of both the atom and the molecule
to the cavity (see also Fig. \ref{Fig.1}). As seen in Fig. \ref{Fig.2},
a nearly periodic energy exchange takes place. Of course, the photonic
mode is also populated and this population is seen to follow the molecular
population. Having all the individual populations of eq. (\ref{eq4})
at hand, we can see in agreement with the expectation from the Hamiltonian
in eq. (\ref{eq:hfull4}) that once the state $\mathrm{\left|G,A_{3},0\right\rangle }$
is populated, it evolves to the state $\mathrm{\left|G,A_{2},1\right\rangle }$,
to which it is directly coupled via the cavity and a photon appears
in the cavity mode allowing now for further population transfer to
the molecular excited state $\left|E,A_{2},0\right\rangle $. It can
be seen that the molecular excited-state population is in phase with
the cavity mode one, while it is opposite to the atomic excited-state
population. In the following we drop the notion excited-state for brevity.
Note, however, that the population of the photonic mode is much smaller
than that of atom and molecule. 

At the lowest displayed coupling value of $\mathrm{g=0.0022\,\omega_{c}}$
(see in Fig. \ref{Fig.2}a), the population stays almost only at the
atom, but as the coupling is increased, it is slowly transferred to
the molecule. Increasing the coupling further to $\mathrm{g=0.0038\,\omega_{c}}$,
one can observe an almost complete Rabi oscillation of the population
between the atom and molecule. By further increasing the coupling,
the amplitude of the Rabi oscillation gradually decreases and smaller
amounts of populations are transferred to the molecule. For more values
of the coupling strengths we refer to Fig. S2 of the (SI). 

Consulting the adiabatic picture allows one to gain additional insight.
It is known, that the adiabatic or polaritonic surfaces can sensitively
change by changing the coupling strength. Initially, the total population
is in the $\mathrm{\left|G,A_{3},0\right\rangle }$ state, which in
the adiabatic picture can be approximated by the lower polariton state
on the left and the middle polariton state on the right (see in Fig.
\ref{Fig.1}c). The excitation is then transferred from here to the
molecule. In the adiabatic picture, the latter is made up essentially
of the middle polariton state on the left and of the lower one on
the right. When the ground vibrational state of the diabatic potential
of $\left|\mathrm{G,A_{3},0}\right\rangle $ coincides with one of
the vibrational states of the lower polariton surface, the population
can migrate from one object to the other, the fingerprint of which
can be clearly seen at $\mathrm{g=0.0038\,\omega_{c}}$ in Fig. \ref{Fig.2}.
The atomic and molecular population curves cross each other smoothly.
Further increasing the coupling strength, the vibrational states of
the two surfaces shift from each other, resulting in only a partial
population exchange. The corresponding vibrational levels of the two
surfaces are gradually moving closer together, then overlap, and again
are moving away from each other as the coupling strength is increased.
By computing the diabatic vibrational eigenstates of the molecule
in its ground state we find that the adiabatic vibrational level of
the lower polariton well coincides with the $\mathrm{\nu=6}$ vibrational
eigenstate. For a more detailed description of the vibrational analyses,
we refer to the Fig. S3 of the SI. The amplitude of the $\mathrm{\nu=6}$
vibration in the total wave function is at least two orders of magnitude
larger than the amplitude of the other diabatic vibrations for all
of the g values applied, but is the largest for $\mathrm{g=0.0038\,\omega_{c}}$.
At this value of the coupling the corresponding vibrational levels
of the two surfaces overlap almost completely. Starting from a small
value of the coupling strength, the contribution of $\mathrm{\nu=6}$
to the molecular population gradually increases and the Rabi frequency
gradually decreases. At $\mathrm{g=0.0038\,\omega_{c}}$ the amplitude
is at its maximum and the Rabi frequency at its minimum, while increasing
$\mathrm{g}$ further, the the trend is inversed. 

It is important to note that the photonic population is very low throughout
the coupling range studied. The atom and the molecule periodically
share the bulk of the total population, the molecular population increases
as $\mathrm{g}$ grows and then decreases after arriving at $\mathrm{g=0.0038\,\omega_{c}}$.
Interestingly, in spite of the involved behavior of the molecular
population as a function of the coupling strength and also of time,
the cavity mode population is always found to be approximately 10\%
of the molecular population.

What is the impact of cavity losses on the above discussed behavior
of the populations? To that end we repeat the calculations employing
a cavity photon lifetime of $\mathrm{\tau=\mathrm{61\,fs}}$ which
is suitable for a plasmonic nano-cavity \citep{Feist2022}. The results
are collected in Fig. \ref{Fig.3} in analogy to those shown in Fig.
\ref{Fig.2} for a lossless cavity. The most striking difference compared
to the lossless case is that the dynamics decays over the time. But
let us discuss first the similarities. It can be seen that initially
the entire population is on the atom, and basically stays there at
lower couplings. If, however, the coupling strength increases, more
and more populations is transferred to the molecule. The atomic and
molecular populations are now still evolving in opposite phases. Similarly
to the lossless situation, the photonic population is again about
an order of magnitude smaller than the other two, and follows the
evolution of the molecular population. As will be seen the variation
in the magnitude of the photonic population plays a significant role
during the lossy dynamics. 

Increasing the coupling strength, the decay of the total population
(sum of populations in Fig. \ref{Fig.3}) first becomes increasingly
more pronounced.The decay is fastest at $\mathrm{g=0.0038\,\omega_{c}}$
where according to panel c) in Fig. \ref{Fig.3}, the total population
nearly disappears after $\mathrm{t=5\,ps}$. One should be aware of
the fact that $\mathrm{5\,ps}$ is still a much longer time than the
lifetime of the cavity photon. Further increasing the coupling strength,
however, shows that there is a significant change in the decay which
starts to resemble that displayed at lower $\mathrm{g}$ values. The
decay time of the total population dynamics is increasing again. It
is clearly seen from the calculations and Fig. \ref{Fig.3} that,
when the population of the photonic mode is higher, the population
decreases faster. Since this value is the largest at $\mathrm{g=0.0038\,\omega_{c}}$,
it is when the fastest relaxation occurs in the system. As $\mathrm{g}$
is increased further, the ground vibrational state of $\mathrm{\left|G,A_{3},0\right\rangle }$
and the corresponding vibrational state of the lower polariton starts
to shift away from each other resulting, as discussed above, in less
molecular population. Correspondingly, the photonic population will
decrease as well, and the decay of the total population becomes slower.
It can be deduced from eq. (\ref{eq:hfull4}) that the gradient of
the total population is proportional to the magnitude of the photonic
population. For more results on additional coupling strengths we refer
to Fig. S4 of the SI. 

The evolution of the various populations has been discussed above
for a lossless cavity and for a cavity photon lifetime of $\mathrm{\tau=61\,fs}$.
We have performed additional computations for cavities of even shorter
lifetimes ($\mathrm{\tau=41\,fs}$ and $\mathrm{\tau=24\,fs}$), see
Fig. S5 and Fig.S6 in the SI. Clearly, the decay of the populations
is faster, but the overall behavior is qualitatively similar to that
shown in Fig. \ref{Fig.3}. 

This letter discusses a dynamical process induced by exciting an atom
in a cavity. The composite system of a molecule and an atom confined
in a microscopic cavity exhibits three polaritonic states which are
hybrid atom-molecule-photon states. Under the conditions discussed,
an external laser source used to pump a certain energy level of the
atom launches, due to the interaction of the atom and the molecule
to the cavity mode, interesting dynamics in the molecule. The initial
excitation of the atom initiates a back-and-forth energy flow between
the different components of the hybrid system which has been analyzed.
Among others, it has been found that the cavity mode is only little
populated and plays the role of a mediator between the atom and the
molecule. 

As demonstrated, the emerging dynamics is long-lived, even in the
case of strong losses of the cavity photon. This makes the dynamics
amenable to experimental observations. To better assess the impact
of the atom on the duration of the dynamics, we repeated the calculations
which lead to Fig. \ref{Fig.3} without the presence of the atom.
To that end the laser is used to pump the molecule directly, i.e.,
the same laser parameters are used (see eq. (\ref{eq4})) except that
the laser frequency is now $\mathrm{1.968\,eV}$ fitting to the energy
difference between the $\mathrm{X^{1}\varSigma_{g}^{+}}$ and $\mathrm{A^{1}\varSigma_{u}^{+}}$
molecular states. The results analogous to those in Fig. \ref{Fig.3}
are displayed in Fig. \ref{Fig.4}. It is eye catching that the dynamics
in the absence of the atom dies off much faster.

We mention that one has plenty of possibilities for choosing atoms
with an appropriate energy gap between two states coupled by a cavity.
Consequently, one can employ various light sources of interest. Here,
we used a Ne atom and a laser frequency of about $\mathrm{20\,eV}$,
but one can also use other atoms like Rb which has many low-lying
energy levels and various light sources of low-energy photons in the
range where the molecule has no accessible energy levels or even of
very high energy photons in the X-ray regime where the atoms have
their core-level excitations. In the latter case, the lifetime of
the levels due to the Auger effect have to be taken into account making
the dynamics more involved. In principle, one can even choose to excite
the atom with a light source which may also excite molecular levels.
In this way, the interplay of molecular levels that do not couple
to the cavity with levels which do can be investigated as well. 

As a last point we stress that it is favorable to have several or
even many atoms in the cavity. Then, the atomic level pumped by the
laser is an atomic polariton which can be excited efficiently by a
weak laser as the transition dipole of the upper polariton is much
larger than that of the individual atom. In this way one can provide
high atomic populations and still employ low intensity pulses. As
the energy of the atomic polariton varies with the number of atoms,
one has an additional tool to choose the energy of the light source.
As usual, having several or many molecules enhances the coupling to
the cavity and makes the interplay and dynamics richer.\vspace*{-3mm}

\begin{acknowledgments}
The authors are indebted to NKFIH for funding (Grant No K146096).
Financial support by the Deutsche Forschungsgemeinschaft (DFG) (Grant
No. CE 10/56-1) is gratefully acknowledged. A.C. is grateful for the
support of the Janos Bolyai Research Scholarship (No. BO/ 00474/22/11)
of the Hungarian Academy of Sciences. 
\end{acknowledgments}

\bibliographystyle{pccp}
\bibliography{energy_transfer}

\end{document}